\documentclass[12pt]{article}
\usepackage{latexsym,slashed,cite,amssymb}
\newcommand{\ft}[2]{{\textstyle\frac{#1}{#2}}}
\def\rmi{{\rm i}}
\def\rmd{{\rm d}}
\newcommand{\hc}{{\rm h.c.}}
\newcommand{\dr}{\raise.3ex\hbox{$\stackrel{\leftarrow}{\delta }$}{}}
\newcommand{\dl}{\raise.3ex\hbox{$\stackrel{\rightarrow}{\delta }$}{}}

\newcommand{\SO}{\mathop{\rm SO}}
\newcommand{\U}{\mathop{\rm {}U}}

\newcommand{\Gl}{\mathop{\rm {}G}\ell }

\csname @addtoreset\endcsname{equation}{section}
\begin{document}

\begin{titlepage}
\begin{flushright}
Imperial/TP/02-03/27 \\
KUL-TF-03/15\\
hep-th/0306244
\end{flushright}
\vspace{.5cm}
\begin{center}
\baselineskip=16pt
{\LARGE    Wess--Zumino sigma models \\[2mm]
with non-K{\"a}hlerian geometry 
}\\
\vfill
{\large Kellogg S. Stelle$^{1}$} \\ \smallskip and \\ \smallskip
{\large Antoine Van Proeyen$^2$} \\
\vfill
{\small $^1$ The Blackett Laboratory, Imperial College London} \\
 Prince Consort Road, London SW7 2AZ, U.K.
\\ \vspace{6pt}
$^2$ Instituut voor Theoretische Fysica - Katholieke Universiteit Leuven
\\
Celestijnenlaan 200D B--3001 Leuven, Belgium
\end{center}
\vfill
\begin{center}
{\bf Abstract}
\end{center}
{\small Supersymmetry of the Wess--Zumino ($N=1$, $D=4$) multiplet allows
field equations that determine a larger class of geometries than the
familiar K{\"a}hler manifolds, in which covariantly holomorphic vectors
rather than a scalar superpotential determine the forces. Indeed,
relaxing the requirement that the field equations be derivable from an
action leads to \textit{complex flat geometry}. The Batalin--Vilkovisky
formalism is used to show that if one requires that the field equations
be derivable from an action, we once again recover the restriction to
K{\"a}hler geometry, with forces derived from a scalar superpotential.

}\vspace{2mm} \vfill \hrule width 3.cm \vspace{2mm}{\footnotesize
\noindent E-mail: \texttt{k.stelle@imperial.ac.uk,
Antoine.VanProeyen@fys.kuleuven.ac.be } }
\end{titlepage}
\section{Introduction}

One of the basic properties of supersymmetric theories is their relation
to the geometry of the target manifold. The prime example of
supersymmetry is the Wess--Zumino multiplet~\cite{Wess:1974kz}. It was
shown early on that in interacting theories of this supermultiplet the
kinetic terms for the scalar fields in the action determine K{\"a}hler
manifolds~\cite{Zumino:1979et}. The interactions are determined by a
scalar holomorphic superpotential $W(z)$, where $z$ represents the
complex scalars of the Wess--Zumino multiplets.

Despite the fact that this multiplet has since then been presented in
many courses in the subject, we discover in this paper that it still
contains some surprises. Relaxing the requirement that the field
equations be derivable from an action, we obtain two generalizations.
First, the kinetic terms of the field equations allow a \textit{complex
flat} geometry, which is a larger class of geometries than the usual
K{\"a}hler geometries. Secondly, the forces are derived from covariantly
holomorphic vectors on the target manifold, instead of from a scalar
superpotential. The `complex flat' condition on the geometry can be seen
as an integrability condition for the existence of these covariantly
holomorphic vectors.

Admittedly, a common starting point for passage to a quantum field theory
is the classical action. However, consistent sets of classical field
equations that are not comprehensively derivable from an action are
well-known. For example, type IIB string theory contains a self-dual
5-form field strength, whose self-duality properties are normally stated
simply as a field equation. More generally, the effective field equations
for the massless modes of string theories arise directly from beta
function conditions. The classical action is not a primary construct in
this derivation.

In this paper, we wish to distinguish between requirements on the
geometry originating strictly from the supersymmetry algebra, and those
that follow from the existence of an action. Examples of geometries wider
than those expected from action formulations are known in the context of
$N=2$ hypermultiplets~\cite{VanProeyen:2001wr,Bergshoeff:2002qk}. A
natural question about such examples is whether their special structure
relates to the extended supersymmetry, or whether they constitute also
exceptions to the expected geometry even of $N=1$ supersymmetry. Here, we
find that, indeed, there exists a wider class of $N=1$, $D=4$
Wess--Zumino sigma models than the standard K{\"a}hlerian ones. This class is
determined purely by the supersymmetry algebra. We find that one can even
allow geometries that are non-Riemannian, since only the affine
connection appears in the supersymmetry transformations and field
equations. The allowed geometries are \textit{complex
flat}~\cite{Joyce:1995}. Moreover, allowed potentials can be derived from
covariantly holomorphic vector functions of the scalar fields (which we
shall refer to as \textit{vector potentials}).

Imposing also the requirement that the field equations be derivable from
an action leads one back to the standard formulation in terms of K{\"a}hler
manifolds and with a scalar superpotential. We use the
Batalin--Vilkovisky (BV)
formalism~\cite{Batalin:1983jr,Henneaux:1990jq,Gomis:1995he} to control
the off-shell non-closure of the algebra. The sigma model metric makes
its appearance here as the matrix that relates the non-closure functional
to the variation of the action. The BV consistency conditions then
require this metric to be K{\"a}hlerian. Using the metric to lower the index
on the (contravariant) vector potential, BV consistency then requires the
curl of the resulting (covariant) vector potential to vanish. Hence, the
standard superpotential makes its appearance by solving the vanishing
curl condition.

The previously known rigid $N=2$ examples turn out to be specializations
of these complex flat geometries. Specifically, the $N=2$ hypermultiplet
geometries are hypercomplex, which bear a similar relation to the
hyperk{\"a}hler ones as the complex flat geometries bear to the K{\"a}hler ones.
Initially, it was suspected~\cite{VanProeyen:2001wr,Bergshoeff:2002qk}
that the origin of the generalized $N=2$ hypermultiplet geometries lay in
the absence of an off-shell formulation for these models. However, in
this paper we see that these generalizations appear owing to the wider
possibilities that exist for auxiliary field equations when they are not
required to be derivable from an action.

In section~\ref{ss:closure} we show how the closure of the supersymmetry
algebra on-shell restricts the allowed field equations. The result, as
shown in section~\ref{ss:complexflat}, is that the target-space manifold
must have a complex flat geometry. In section~\ref{ss:superspace}, we
show all of this again in superspace. In section~\ref{ss:BVaction} we
show how one recovers the usual K{\"a}hler geometry with a superpotential
once one requires the field equations to be derivable from an action. We
use the BV formalism to give a general derivation that does not depend on
any assumption of an off-shell supersymmetric or superspace action. In
the concluding section~\ref{ss:conclusion}, we discuss possible
extensions of these results to supergravity couplings.

\section{Closure conditions constrain the geometry}\label{ss:closure}

In order to make as few restrictive assumptions as possible, we will work
with explicit component physical fields only, \textit{i.e.}\ without
auxiliary fields. We will investigate how the $N=1$, $D=4$ supersymmetry
algebra can be realized on chiral multiplets, with closure following
solely from \emph{dynamical} constraints on the physical fields.
Specifically, for 2 supersymmetry transformations with rigid parameters
$\epsilon _1$ and $\epsilon _2$, we require the commutator
\begin{equation}
  \left[ \delta (\epsilon _1), \delta (\epsilon _2)\right] = \bar \epsilon
  _2 \gamma ^\mu \epsilon _1 \partial _\mu =
   \left( \bar \epsilon _{2L} \gamma ^\mu \epsilon _{1R} +\bar \epsilon _{2R} \gamma ^\mu \epsilon _{1L}\right)
\partial _\mu\,,
 \label{susyalg}
\end{equation}
to be realized on all fields, up to dynamical equations of motion.

By (left-handed) chiral multiplets we mean the following. These are
multiplets of complex scalars $z^a$, where $a$ is the label of a
particular multiplet, together with their spinor superpartners $\zeta
^a_L$. The left-handed spinor $\zeta ^a_L$ is defined as normal to
satisfy
\begin{equation}
  \zeta^a _L=\ft12(1+\gamma _5)\zeta^a_L \,,
 \label{notationzetaL}
\end{equation}
where $(\gamma_5)^2=1$. The right-handed spinors are charge conjugates of
the corresponding left-handed ones,
\begin{equation}
  \zeta^{\bar a} _R= \left( \zeta^a _L\right) ^*= \ft12(1-\gamma _5)\zeta _R^{\bar
  a}\,.
 \label{zetaR}
\end{equation}

Without further ado, we assume that the target manifold of the $z^a$
(with complex conjugates $z^{\bar a}$) is a complex manifold, with
complex structure
\begin{equation}
  J_a{}^b=\rmi \delta_a{}^b\,,\qquad J_{\bar a}{}^{\bar b}=-\rmi \delta_{\bar a}{}^{\bar b}\,,\qquad
  J_a{}^{\bar b}=J_{\bar a}{}^b=0\,.
 \label{ConstComplexstructures}
\end{equation}
The definition of a chiral multiplet is now that $z^a$ transforms only
with a (left-handed) chiral supersymmetry parameter:
\begin{equation}
  \delta (\epsilon) z^a= \bar \epsilon_L \zeta_L ^a\,, \qquad \mbox{where}\qquad
  \bar \epsilon _L=\ft12\bar \epsilon _L(1+\gamma _5)\,.
 \label{delQz}
\end{equation}
Note that the Lorentz conjugate $\bar\epsilon_L$ used here may also be
given as $\epsilon_R^\dag\gamma^0$. Accordingly, $\bar\zeta^a_L$ has the
same holomorphic index type as $\zeta^a_L$.

Now we look for the most general transformation of the fermions. This can
be parametrized as\footnote{See this approach already
in~\cite[section~3.1]{VanProeyen:1983wk}.}
\begin{equation}
  \delta (\epsilon) \zeta_L ^a= \gamma ^\mu \epsilon _R X_\mu^a + \gamma ^{\mu \nu }\epsilon
  _L T^a_{\mu \nu } + h^a \epsilon _L\,.
 \label{delQzeta}
\end{equation}
Imposing the supersymmetry algebra~(\ref{susyalg}) on the scalars leads
to
\begin{equation}
  X_\mu^a =\partial _\mu z^a\,, \qquad T^a_{\mu \nu }=0\,,
 \label{impossusyz}
\end{equation}
while $h^a$ remains an arbitrary complex function of the scalars and
fermions. Thus we obtain
\begin{equation}
  \delta (\epsilon) \zeta_L ^a= \slashed{\partial }z^a \epsilon _R
+ h^a \epsilon _L\,.
 \label{delQzetaSol}
\end{equation}

Checking the supersymmetry algebra on the spinors, we find that closure
(allowing only constraints that are dynamical on the physical fields $z$
and $\zeta $) requires that $\delta (\epsilon) h$ does not contain
$\epsilon _L$:
\begin{equation}
  \delta h^a =\bar \epsilon _R\lambda _R^a \,,
 \label{delh}
\end{equation}
where $\lambda _R^a$ is a so-far undetermined right handed spinor. The
resulting algebra for the fermions is
\begin{equation}
    \left[ \delta (\epsilon _1), \delta (\epsilon _2)\right]\zeta_L ^a =
    \left(   \partial _\mu\zeta_L ^a
  -\ft12 \gamma _\mu I_R^a\right)  \bar \epsilon
  _2 \gamma ^\mu \epsilon _1\,,
 \label{nonclosFerm}
\end{equation}
where
\begin{equation}
  I_R^a=\slashed{\partial }\zeta_L^a -\lambda_R^a
 \label{nonclos}
\end{equation}
is the non-closure functional.

We require that $h^a$ be given in terms of the physical fields. (Normally
$h^a$ would be the on-shell value of the auxiliary field as determined by
an action, but we are not making such assumptions here). We assume only
that the $h^a$ are scalars with respect to the Lorentz group. We may
expand in powers of fermions; dimensional arguments restrict this to
terms at most quadratic in the fermions:
\begin{equation}
  h^a =W^a(z,\bar z)
  + \tilde \Gamma _{\bar b\bar c }{}^a(z,\bar z)\bar \zeta
  _R^{\bar b} \zeta _R^{\bar c }
  +\ft12\Gamma  _{bc}{}^a(z,\bar z) \bar\zeta _L^b \zeta _L^c \,,
 \label{htrial}
\end{equation}
where $W^a$, $\tilde \Gamma _{\bar b\bar c }{}^a$ and $\Gamma  _{bc}{}^a$
are so-far undetermined functions of the scalars ($z^a,\bar z^{\bar a}$).
The conditions for the transformation of $h^a$ not to contain $\epsilon
_L$ are
\begin{eqnarray}
 \nabla _b W^a  & \equiv  & \partial _bW^a + \Gamma  _{bc}{}^a W^c =0 \,,\label{covderW}\\
 \tilde \Gamma _{\bar b \bar c}{} ^a & = & 0 \,,\label{mixedGamma}\\
 R_{abc }{}^d &\equiv & 2\partial _{[a}\Gamma _{b ]c}{}^d +2\Gamma _{e[a}{}^d\Gamma _{b]c}{}^e=0\,.\label{flatR}
\end{eqnarray}
Thus, the final expression for $h^a$ is
\begin{equation}
  h^a =W^a(z,\bar z)
  + \ft12\Gamma  _{bc}{}^a(z,\bar z) \bar\zeta _L^b \zeta _L^c \,,
 \label{hWGamma}
\end{equation}
and the non-closure functional~(\ref{nonclos}) is
\begin{equation}
  I_R^a=\slashed{\nabla}\zeta_L^a - \zeta ^{\bar b}_{R}\partial _{\bar
  b}W^a -\ft12\zeta _R^{\bar d} \bar\zeta _L^b\zeta _L^c R_{\bar dbc}{}^a\,,
 \label{IRa}
\end{equation}
where
\begin{equation}
  \nabla_\mu\zeta_L^a\equiv\partial _\mu \zeta_L^a +\zeta _L^b \Gamma _{bc}{}^a\partial_\mu
  z^c\,, \qquad R_{\bar dbc}{}^a\equiv  \partial _{\bar d}\Gamma_{bc}{}^a\,.
 \label{DzetaR}
\end{equation}
Setting $I_R^a=0$ gives the fermionic field equation.

The bosonic field equation follows from a supersymmetry variation of the
fermionic one. This is
\begin{eqnarray}
&&\Box z^a - {\bar W}^{\bar b}\partial _{\bar b
   }W^a+\ft12\bar \zeta _R^{\bar b}\zeta _R^{\bar c}\nabla_{\bar b}\partial _{\bar c}W^a\nonumber\\
&&  -\ft12\bar \zeta _L^b\zeta _L^c\bar W^{\bar d}R_{\bar dbc}{}^a
  +\ft14\bar \zeta _R^{\bar e}  \zeta _R^{\bar d}\bar \zeta _L^b\zeta _L^c\nabla _{\bar e}R_{\bar dbc}{}^a
  -\bar \zeta _R^{\bar d}\gamma ^\mu \zeta _L^b\partial _\mu z^c R_{\bar dbc}{}^a
  =0\,.
 \label{fieldeqns}
\end{eqnarray}
The target-space covariant derivative $\nabla _b$ was already introduced
in~(\ref{covderW}) when acting on a target space vector; on more general
tensors like $R_{\bar dbc}{}^a$ it acts according to the standard
covariant rule. From this we already note that $\Gamma $ acts as a
connection. We shall return to the details of this geometry in
section~\ref{ss:complexflat}. The covariant d'Alembertian $\Box$ is
defined using a pulled back covariant derivative $\nabla_\mu $:
\begin{equation}
  \Box z^a=\nabla^\mu \partial _\mu z^a=\partial ^\mu \partial _\mu z^a +\Gamma ^a_{bc}\left(
\partial ^\mu z^b\right) \left( \partial _\mu z^c\right) .
 \label{covbox}
\end{equation}

Then one checks that the supersymmetry transformation of the bosonic
field equation leads to no new conditions, \textit{i.e.}\ it leads only
to equations involving derivatives of the original one. This will become
more clear using superspace notation in the next section.

The interesting result here is that we find the
requirements~(\ref{covderW} -- \ref{flatR}). These lead us to the concept
of `\emph{complex flat}' geometry, as we will show in the next section.

\section{Complex flat geometry}\label{ss:complexflat}

In the previous section we found  a complex geometry, with complex
structures defined by~(\ref{ConstComplexstructures}). It has a
torsionless affine connection with purely holomorphic indices $\Gamma
_{bc}{}^a$ (together with its complex conjugate $\Gamma _{\bar b\bar
c}{}^{\bar a}$). The torsionless condition is the symmetry of the
connection as is manifest from its introduction in~(\ref{htrial}).

The purely holomorphic character of the connection can be interpreted as
the condition for the complex structure~(\ref{ConstComplexstructures}) to
be covariantly constant. This promotes the scalar-field target space from
an almost complex to a complex manifold. Explicitly, the vanishing of the
mixed components implies
\begin{eqnarray}
 &  & \nabla _{\bar c}J_{\bar a}{}^b= -\Gamma_{{\bar c}{\bar a}}{}^dJ_d{}^b
 + \Gamma_{{\bar c}{\bar d}}{}^bJ_{\bar a}{}^{\bar d}=-2\rmi \Gamma_{{\bar c}{\bar a}}{}^b=0\,, \nonumber\\
 & & \nabla _{\bar c}J_{a}{}^{\bar b}= -\Gamma_{{\bar c}a}{}^{\bar d}
 J_{\bar d}{}^{\bar b}
 + \Gamma_{{\bar c}d}{}^{\bar b}J_a{}^d=2\rmi \Gamma_{{\bar c}a}{}^{\bar
 b}=0\,.
 \label{zeroGammamixed}
\end{eqnarray}

The remaining restriction~(\ref{flatR}) that we found on the geometry,
now interpreted as the vanishing of the purely holomorphic components of
the curvature, defines~\cite[Definition~6.1]{Joyce:1995} a
\textit{complex flat} manifold.

Aside from the geometry, our results include a covariantly holomorphic
vector potential $W^a$, see~(\ref{covderW}). This generalizes the
gradient of the scalar superpotential that appears in the standard sigma
models, where one assumes the existence of an action. The integrability
condition for such a covariantly holomorphic vector to exist is precisely
the complex flat condition $R_{abc}{}^d=0$.

Examples of such geometries are the hypercomplex geometries that were
found in the context of $N=2$ supersymmetric field equations without an
action, for a system of hypermultiplets. This follows from
Proposition~7.1 in~\cite{Joyce:1995}, which states that a hypercomplex
manifold is necessarily complex flat with respect of any one of its
complex structures. Therefore the explicit examples based on group
manifolds given in~\cite{Spindel:1988sr,Bergshoeff:2002qk} are therefore
also explicit examples of the complex flat manifolds considered here.

\section{Covariant superspace reformulation} \label{ss:superspace}

We reformulate our results now in two respects. The transformations of
the fermions can be written in a way that is known from the standard
supersymmetric sigma models, and we can also introduce a covariant
superspace formulation. The former is clear from combining the
results~(\ref{delQzetaSol}) and~(\ref{hWGamma}), using a Fierz
transformation to give
\begin{eqnarray}
    \delta (\epsilon) \zeta_L ^a &=& \slashed{\partial }z^a \epsilon _R
+ W^a \epsilon _L + \ft12 \zeta_L^c \bar\zeta_L^b\epsilon_L  \Gamma_{bc}^a \label{covfermionvar}\\
&=& \slashed{\partial }z^a \epsilon _R + W^a \epsilon _L-\zeta_L ^b
\Gamma_{bc}{}^a \delta (\epsilon)z^c\,.
 \label{delQzetaGamma}
\end{eqnarray}
We shall soon see how this can be interpreted as a covariant superspace
transformation.

The superspace formulation of our results uses standard chiral
superfields $\phi ^a$, \textit{i.e.}\ $\bar D_{\dot\alpha }\phi ^a=0$.
The dynamics is defined by imposing an extra condition that is the
superspace version of~(\ref{hWGamma}) together with its transformed
equations~(\ref{IRa}) and~(\ref{fieldeqns}):
\begin{equation}
  \nabla ^\alpha D_\alpha \phi ^a= 4\rmi W^a\,,
 \label{general}
\end{equation}
where $\nabla_\alpha$ is defined similarly to $\nabla_\mu$, \textit{i.e.}
\begin{equation}
\nabla_\alpha\lambda_\beta^a = D_\alpha\lambda_\beta^a + D_\alpha
\phi^b\Gamma_{bc}^a\lambda_\beta^c \,.\label{nablaalphadef}
\end{equation}
Equation~(\ref{covfermionvar}) is now recognizable as the component
version of the covariantized superspace supersymmetry transformation
generated by $(\nabla_\alpha,\bar\nabla_{\dot\alpha})$.

The superspace equation of motion should be covariantly antichiral in
order for the equations following from it by differentiation to involve
only the expected dynamical equations for the physical fermions and
scalars. In particular, we want to avoid the appearance of algebraic
constraints on the physical fields.

In order for the left-hand side to be covariantly antichiral
(\textit{i.e.}\ to be annihilated by $\nabla _\beta $), we require
$\{\nabla _\alpha ,\nabla _\beta \}=0$. This in turn
requires~(\ref{flatR}). For the right-hand side this is accomplished by
having $W^a$ be covariantly holomorphic, \textit{i.e.}~(\ref{covderW}).
Thus, the superspace formulation succinctly summarizes all of the
geometrical conclusions arrived at in section~\ref{ss:closure}.

\section{Requiring an action recovers the standard geometry}\label{ss:BVaction}

We will now see how requiring that equations of motion be derivable from
an action leads one back to K{\"a}hler geometry and forces derived from a
scalar superpotential. The above complex flat geometry, as far as one can
see, is free from the inconsistencies that may accompany equations that
are not derivable from an action. In particular, requiring the geometry
to be complex flat and the vector potential to be covariantly holomorphic
precludes the appearance of hidden constraints in the equations of
motion. It is clear that these equations cannot be derived from a
standard action, however, because a sigma-model action requires a target
space metric, which we have not even defined. In short, the above
geometries are purely affine.

Clearly it is of interest to see what further restrictions on the
geometry arise when one requires the existence of an action. Instead of
following the textbook approach that starts from a superspace
formulation~\cite{Zumino:1979et}, we will try to make as few assumptions
as possible, and will derive the K{\"a}hler geometry from the BV formalism
without specifying a particular classical action.\footnote{The BV
formalism was also used to determine possibilities in supersymmetry and
supergravity in~\cite{Brandt:1997au,Brandt:1998gj}.}

The Batalin--Vilkovisky (BV) or `antifield'
formalism~\cite{Batalin:1983jr,Henneaux:1990jq,Gomis:1995he} is a
convenient bookkeeping device to keep track of the elementary relations
of gauge theories. Here we consider only rigid supersymmetric theories,
but we can nonetheless use a formalism with ghosts (and their antifields)
that are taken to be constants rather than spacetime-dependent
fields\footnote{For more details on the use of rigid symmetries in the BV
formalism, see~\cite{Brandt:1996uv}.}. We introduce ghosts $c^\mu $ for
translations and $c_L$ and $c_R$ for supersymmetry.

Briefly, the BV formalism introduces an antifield $\Phi ^*_A$ for every
field $\Phi ^A$, including the ghosts, and the main task is to ensure the
validity of the master equation for the `extended' action $S_{BV}(\Phi
,\Phi ^*)$:
\begin{equation}
  (S_{BV},S_{BV})= 2S_{BV}\frac{\dr}{\delta  \Phi^A }\, \frac{\dl}
  {\delta  \Phi^*_A}S_{BV}=0\,.
 \label{master}
\end{equation}
Expanding in terms of antifield number (see table below), this extended
action begins with a classical action $S_0(\Phi )$, which we do not
specify. Suppressing integrals $\int \rmd^4 x $, and using the
supersymmetry algebra and physical field transformations and that we
already computed, we can write the BV action as
\begin{eqnarray}
 S_{\rm BV} & = & S_0(z,\bar z,\zeta _L,\zeta _R)+S_1+S_2\,, \nonumber\\
S_1&=&z^*_a c^\mu \partial _\mu z^a +\zeta ^*_{La} ( \partial _\mu \zeta
_L^a)c^\mu+
  z^*_a\bar \zeta _L^ac_L+\bar \zeta _{La}^*\left[ \slashed{\partial }z^ac_R
  +h^a(z,\bar z,\zeta _L)c_L \right]
  +\hc \,,\nonumber\\
 S_2 & = & c^*_\mu \bar c_L\gamma ^\mu c_R + \ft12\bar \zeta ^*_{La}\gamma _\mu \zeta
 ^*_{R\bar b} \bar c_L\gamma ^\mu c_R\, g^{a\bar b}(z,\bar z)\,.
 \label{SBV}
\end{eqnarray}

The last term in $S_2$ involves a new quantity $g^{a\bar b}(z,z^*)$,
which in non-BV language determines how the non-closure functional $I^a$
is proportional to the field equations:
\begin{equation}
  I_R^a =- g^{a\bar b}\frac{\delta S_0}{\delta \bar \zeta _R^{\bar b}}\,.
 \label{IgdS}
\end{equation}
In the check of the master equation this occurs in the term proportional
to $ \bar c_L\gamma _\mu c_R\bar \zeta ^*_L \gamma ^\mu $. It is this
quantity $g^{a\bar b}$ that will become (the inverse of) the metric.

To see why the action~(\ref{SBV}) stops at antifield number~2, we have to
characterize all the fields and antifields in a 3-fold way, listing
dimension, ghost number and antifield number for each field and antifield
of the theory. In order for the antibracket operation $(\cdot\,, \cdot)$
to be dimensionality consistent, the dimensions of any field and its
antifield should add up to the same number for all pairs. We choose this
to be~1. Note that, owing to ghost number conservation, there is an
arbitrariness in assigning dimensions. We choose the assignments in
table~\ref{tbl:dimghafn},
\begin{table}[htbp]
  \caption{\it Dimensions, ghost and antifield numbers of all independent
  fields and their antifields, the action and the auxiliary fields}\label{tbl:dimghafn}
\begin{center}
$  \begin{array}{|c|ccc|c|ccc|} \hline\hline
 \mbox{field} & \mbox{dim} & \mbox{gh} & \mbox{afn} & \mbox{antifield} & \mbox{dim} & \mbox{gh} & \mbox{afn} \\
\hline
 z^a & 0 & 0 & 0 & z_a^* & 1 & -1 & 1 \\
 \zeta^a  & \ft12 & 0 & 0 & \zeta_a^* & \ft12 & -1 & 1 \\
 c^\mu  & 0 & 1 & 0 & c^*_\mu  & 1 & -2 & 2 \\
 c & \ft12 & 1 & 0 & c^* & \ft12 & -2 & 2 \\ \hline\hline
 h^a & 1 & 0 & 0 & & &  &  \\
 S & 2 & 0 &  & &  &  &  \\
   \hline\hline
\end{array}$
\end{center}
\end{table}
which have the advantage that no fields are of negative dimension. The
action has dimension~2. It has no definite antifield number, and the
expansion in~(\ref{SBV}) is according to antifield number. The
dimensional arguments can be used to see that $g^{a\bar b}$ can only be a
function of the scalars.

One can now easily establish, owing to the absence of objects with
negative dimension, that if there were further terms in the BV action,
they would have to involve the $c^\mu $ ghosts. We do not expect such
terms because the algebra of translations is well behaved, but it can
also easily be checked that even if such terms were to occur, they would
not spoil the arguments below.\bigskip

We now consider the terms in $(S,S)$ that are quadratic in the $\zeta ^*$
antifield and are cubic in the supersymmetry ghosts. The terms quadratic
in $c_L$ and linear in $c_R$ are
\begin{equation}
  (S,S)=\bar \zeta ^*_{La}\gamma _\mu \zeta
 ^*_{R\bar b} \bar c_L\gamma ^\mu c_R \partial _cg^{a\bar b}(z,\bar z)\bar \zeta
 _L^c c_L+ \bar \zeta _{La}^* c_L\left[ h^a(z,\bar z,\zeta _L)\frac{\dr}{\delta \zeta
 _L^d} \right] \gamma _\mu \zeta
 ^*_{R\bar b} \bar c_L\gamma ^\mu c_R g^{d\bar b}(z,\bar z)\,.
 \label{SSzeta*zeta*}
\end{equation}
The vanishing of this expression, using~(\ref{hWGamma}), leads after a
few Fierz transformations to
\begin{equation}
 (S,S)=-\ft18 \bar \zeta ^*_{La}\gamma _\mu \zeta
 ^*_{R\bar b}\bar c_L\gamma_{\rho \sigma }c_L \bar \zeta _L^c\gamma^{\rho \sigma }\gamma
 ^\mu c_R\left[
 \partial _cg^{a\bar b}(z,\bar z) +\Gamma  _{dc}{}^a(z,\bar z)  g^{d\bar b}(z,\bar z)\right].
 \label{SSDg0}
\end{equation}

This proves that the $g^{a\bar b}$ is covariantly constant for the affine
connection that we introduced. We may now define the inverse of this
$g^{a\bar b}$ as $g_{a\bar b}$:
\begin{equation}
 g_{a\bar b}g^{c\bar b} =\delta _a{}^c\,.
 \label{gginv}
\end{equation}
Then we have
\begin{equation}
\partial _c  g_{a\bar b}-\Gamma _{ca}{}^d g_{d\bar b}=0\,.
 \label{Dg0}
\end{equation}
The symmetric part in $(ca)$ of this equation determines that $\Gamma $
is the Levi-Civita connection of $g$:
\begin{equation}
  \Gamma _{ab}{}^c = g^{c\bar d}\partial _{(a}g_{b)\bar d}\,.
 \label{GammaLC}
\end{equation}
The antisymmetric part says that $g$ is a K{\"a}hler metric:
\begin{equation}
  \partial _{[c}  g_{a]\bar b}=0\,.
 \label{gKahler}
\end{equation}

We derived~(\ref{IgdS}) as a consequence of the BV equations. We can
write this now as (introducing spinor indices $\alpha $ now to be
explicit)
\begin{equation}
 g_{a\bar b} I_{R\alpha }^a  =- \frac{\dr S_0}{\partial   \zeta _R^{\alpha  \bar b}}\,.
 \label{dS0}
\end{equation}
We have not committed ourselves so far to any particular expression for
the classical action $S_0$. But if any $S_0$ should exist, we obtain here
an integrability condition
\begin{equation}
 g_{a\bar b} \frac{\dr }{\partial   \zeta _R^{\beta   \bar c}}
 I_{R\alpha }^a + (\bar b\alpha \leftrightarrow \bar c\beta ) =0\,.
 \label{integrability}
\end{equation}
Using the explicit expression~(\ref{IRa}) and working to lowest order in
the fermions, gives, with ${\cal C}_{ \alpha\beta }$ the (antisymmetric)
charge conjugation matrix,
\begin{equation}
  g_{a\bar b} \partial _{\bar c}W^a {\cal C}_{ \alpha\beta }
  + (\bar b\alpha \leftrightarrow \bar c\beta ) =0\qquad  \rightarrow\qquad
  g_{a[\bar b} \partial _{\bar c]}W^a =0\,.
 \label{integrW}
\end{equation}
As we know meanwhile that $g$ is covariantly constant, we derive from
this that
\begin{equation}
  \partial _{[\bar c}W_{\bar b]}=0\,, \qquad \mbox{with}\qquad  W_{\bar b}= g_{a\bar b}W^a
  \,.
 \label{dW}
\end{equation}
Thus, $W_{\bar a}$ is (locally) derivable from a `prepotential' $W$:
\begin{equation}
  W_{\bar a}=\partial _{\bar a} \bar W\,.
 \label{prepotential}
\end{equation}
The condition~(\ref{covderW}) then implies that $\bar W$ is
antiholomorphic. Thus we have recovered all the usual features of actions
for the Wess-Zumino chiral multiplet.

\section{Conclusion and discussion} \label{ss:conclusion}

We have found that imposing the supersymmetry algebra on the fields of
the WZ chiral multiplet leads to `complex flat geometry' rather than
K{\"a}hler geometry, \textit{i.e.}\ to connections that have only holomorphic
indices (together with their complex conjugates) and that satisfy
\begin{equation}
R_{abc }{}^d \equiv  2\partial _{[a}\Gamma _{b ]c}{}^d+2\Gamma
_{e[a}{}^d\Gamma _{b]c}{}^e =0\,.
 \label{complexflat}
\end{equation}
The potential terms depend on arbitrary covariantly holomorphic vectors
$W^a(z,\bar z)$:
\begin{equation}
  \nabla _b W^a(z,\bar z)=0\,.
 \label{DW0}
\end{equation}

If we impose the further requirement that an action invariant under these
transformations exists, then we reobtain the usual extra condition that
the connection is the Levi-Civita connection of a K{\"a}hler metric, and that
the potential is derivable from a scalar holomorphic prepotential.

Clearly, a major question about the physical acceptability of the wider
class of complex-flat sigma models that we have discussed here concerns
their possible quantization. Without an action, standard path-integral
approaches do not apply. But there are other ways to quantize using the
equations of motion directly, \textit{e.g.}\ Schwinger-Dyson
quantization. The effective field equations following from string theory
also arise directly from beta function conditions, without the classical
action playing a direct r{\^o}le. Admittedly, Zamolodchikov's $C$-theorem
does give the central charge functional a r{\^o}le as an effective action, so
the conformal field theory implications of the target space geometries
that we have studied would have to be carefully re-examined.

A related point involves the use of dimensional considerations. In our BV
examination of the consequences of assuming that an action exists, we
restricted the allowed terms to those that are dimensionally consistent
with standard second-order scalar kinetic terms. After quantization,
these restrictions could be relaxed, potentially allowing the BV
discussion to reach a more general conclusion than the standard K{\"a}hler
geometry with a scalar superpotential that we reobtained for theories
derivable from an action.

To summarize the allowed manifolds without and with the assumption that
there exists a preserved metric, we have the situation as shown in Table
\ref{tbl:KahlerlikeMan}. A general curvature tensor with non-zero
components $R_{\bar abc}{}^d$ (and complex conjugates), allows a general
holonomy $\Gl(n,\mathbb{C})$. When one requires an action, the holonomy
group becomes a subgroup of $\SO(2n)$. The intersection of
$\Gl(n,\mathbb{C})$ with $\SO(2n)$ is $\U(n)$.

\begin{table}[ht]
 \begin{center}
  \begin{tabular}{|c|c|}\hline
    no preserved  metric & with a preserved metric \\ \hline\hline
    complex flat & K{\"a}hler \\
    $\Gl(n,\mathbb{C})$ & $\U(n)$ \\ \hline
  \end{tabular}
 \caption{\it Manifolds for chiral multiplet couplings in $D=4$, $N=1$ supersymmetry.
  The corresponding holonomy groups are as indicated, assuming positive
definite kinetic energies for theories derivable from an
action.}\label{tbl:KahlerlikeMan}
\end{center}
\end{table}

We may also think about how this framework can be embedded into local
supersymmetry, \textit{i.e.}\ in the presence of supergravity. This can
be approached using superconformal tensor calculus. One first imposes
conformal symmetry on the geometry and includes at the same time an extra
chiral multiplet, called the compensating multiplet. The conformal
symmetry requires the existence of a closed homothetic Killing vector on
this extended scalar space. The superconformal symmetry is then gauged
using the Weyl multiplet. The latter involves the graviton, the gravitino
and a $\U(1)$ gauge field. Indeed, the superconformal group contains such a
$\U(1)$, which is the R-symmetry group. Then the superfluous gauge
symmetries are removed by gauge fixing. This removes one complex
scalar (corresponding to dilatation and $\U(1)$ gauge fixings) and one
fermion (corresponding to the special supersymmetry present in the
superconformal group). These steps are similar to what has been done for
hypermultiplets in~\cite{Bergshoeff:2002qk} and in a forthcoming
paper~\cite{Bergshoeff:2003yy}.

In general, supergravity coupling leads to a nontrivial factor in the holonomy group
corresponding to the R-symmetry. In the superconformal framework this comes from the
$\U(1)$ mentioned above. A $\U(1)$ holonomy component can also be
non-trivial for purely flat space theories without supergravity coupling,
but the couplings in the fermionic sector of the theory would then be
different from those arising from supergravity coupling. One thing that is
known to change in the presence of supergravity coupling is
an integrality condition on the scalar target manifold. When an action exists, the
K{\"a}hler gauge transformation needs to be accompanied by a $\U(1)$ super-Weyl
transformation upon coupling to supergravity. For theories with an action, the presence
of this $\U(1)$ factor, taken together with
global requirements, leads to a restriction from K{\"a}hler manifolds to Hodge-K{\"a}hler
manifolds~\cite{Witten:1982hu}. Such manifolds have integral periods for the K\"ahler
form. The analogous structure without the assumption that an action exists
remains to be clarified.

We find it amusing that the fundamental multiplet of supersymmetry has
reserved some surprises for us, even 29 years after its introduction.

\medskip
\section*{Acknowledgments and Dedication.}

\noindent We would like to acknowledge stimulating discussions on this
work with Ian Kogan, only two weeks before his tragic death. Ian was a
wonderful person and an ever-enthusiastic colleague and we shall deeply
miss him. This article is dedicated to his memory.
\medskip

This work was supported in part by the European Community's Human
Potential Programme under contract HPRN-CT-2000-00131 Quantum Spacetime.
The work of A.V.P. was supported in part by the Federal Office for Scientific,
Technical and Cultural Affairs through the Inter-university Attraction
Pole P5/27. The work of K.S.S. was supported in part by PPARC under SPG grant
PPA/G/S/1998/00613.
\newpage
\providecommand{\href}[2]{#2}\begingroup\raggedright\endgroup

\end{document}